# Data-informed healthcare service design for multiple long-term conditions using online patient stories

Ji Han, Marta Staff, Saeema Ahmed-Kristensen

The University of Exeter, United Kingdom

**Abstract:**

Conventional service design methods are valuable for improving healthcare experience, but are limited in scale and information capture. Based on a constructed database of 2,320 stories from patients and carers with multiple long-term conditions (MLTC), this paper shows how real-life experiences can be used to inform healthcare service redesign. By combining the richness of qualitative insight with the breadth and representativeness of large-scale data, it identifies "Continuity of care", "Care coordination", and "Temporal - Access to services" as the priority redesign opportunities for MLTC.

## 1. Introduction

Service design often uses methodologies such as user journeys or touch points to understand how to improve the design of services. These methods typically map an individual's journey to identify pain points that can be improved or use interviews to develop personas. They are commonplace amongst service designers in industry and offer rich data, but are time-consuming as they require engagement with end users. Increasingly, data, whether collected from warranty reports or Internet of Things (IoT) sensor technology in devices, are being used to complement these approaches by offering scalable insights into user experiences. When it comes to redesigning healthcare services, service design can uncover pain points and improvement opportunities that inform more coherent, patient-centred models of care (Rego et al., 2022). However, the decision-making remains largely governed by public authorities and still relies heavily on engagement with patients and carers with lived experience. Although the data collected often is of a rich qualitative nature, it typically reflects the experiences of a small number of participants.

We propose to develop a database that enables complementary quantitative data, which enables a far greater representation of people. We argue this is more inclusive and diverse, and can be used to understand areas that require further user stories. Our case study data is within the healthcare of the UK's National Health Service, and particularly for people with multiple long-term conditions (MLTC), which is defined as the coexistence of two or more physical, mental, or infectious conditions lasting for at least 12 months (Thompson et al., 2025). This is an escalating challenge for healthcare systems driven, at least in part, by increased longevity and the rising prevalence of chronic health conditions (Dambha-Miller et al., 2022; Thompson et al., 2025). In the UK, it is estimated that one in four adults is living with two or more health conditions (Department of Health & Social Care, 2023). Importantly, living with MLTC is not limited to older adults. Since 2011, more than half of the new MLTC cases each year have been among people under 50 years of age, and the median age at which complex MLTC develops is below 60 (Head et al., 2021). However, people with MLTC, as well as their health and care professionals (also known as "carers" in this paper), often experience healthcare services that are inefficient, fragmented, and poorly coordinated, resulting in higher treatment costs and reduced quality of life. This growing complexity presents a significant challenge for the current healthcare systems, which are designed primarily around single-disease models of care.

Conventional healthcare systems are typically structured by clinical speciality, resulting in limited continuity and coordination across different services. For people with MLTC, this can result in repetitive assessments, conflicting clinical advice, and inefficient use of resources. Addressing these issues requires redesigning the healthcare services to better reflect patients' lived experiences. However, traditional methods are often small in scale and may not fully capture the diversity and longitudinal nature of experiences among MLTC patients and their carers. Thus, there is a need to combine the richness of experiential data with the representativeness of large-scale, real-world evidence to enable more comprehensive, data-informed service design directions that can address the complexity of MLTC care at both individual and system levels.

To address this, this paper aims to identify and understand priority challenges in healthcare services for MLTC patients and their carers, using large-scale online real-life stories and experiences to inform MLTC healthcare service design. It demonstrates a data-informed approach that complements traditional methods and supports a broader, more inclusive perspective.

2. Theoretical background

Designing effective healthcare services for people with MLTC requires a clear understanding and definition of concepts of continuity of care and care coordination, both of which underpin the delivery of high-quality, patient-centred services. Continuity of care refers to "how an individual's health care is connected over time" (Guthrie et al., 2008). It is particularly important for patients with complex health conditions, where the focus is on maintaining the quality of care over time by improving healthcare services (Gulliford et al., 2006). For patients and their carers, continuity means that care providers know their history and agree on a management plan, while they can also receive care from someone familiar. For care providers, continuity indicates that they will have sufficient information and knowledge to deliver appropriate care, which is recognised and followed by others (Haggerty et al., 2003). It is evidenced in practice that increased continuity of care is positively associated with patient outcomes and satisfaction (Van Walraven et al., 2010). Three types of continuity of care are commonly recognised: informational, management (also known as longitudinal), and relational (Baker et al., 2007; Haggerty et al., 2003). Informational continuity of care refers to the use of information, such as past events and personal circumstances, to deliver care suitable for the individual. Management continuity refers to a consistent and coherent approach to managing a health condition that adapts to the changing needs of the patient over time. Relational continuity concerns the ongoing therapeutic relationship maintained between patients and one or more care providers.

Care coordination refers to the intentional organisation of patient care activities between two or more participants, including patients and care providers, to ensure appropriate healthcare services are delivered (Quinonez et al., 2022). It is considered a key component for delivering high-quality, high-value, and patient-centred healthcare (Schultz & McDonald, 2014), and has been linked to reduced hospital stays, lower inpatient costs, decreased use of inpatient services, and greater patient satisfaction (Berry et al., 2013). It is widely acknowledged that poor care coordination decreases the quality of care and increases negative outcomes, including medication errors and avoidable hospital visits (Schultz et al., 2013). Care coordination encompasses both internal processes, such as medication communication and discharge planning (Figueroa et al., 2018), and external processes, such as coordination between hospitalists and primary care providers (Jones et al., 2015).

It is shown that maintaining continuity of care for MLTC is more cost-effective and also associated with fewer complications (Hill et al., 2014). A lack of continuity of care contributes to healthcare system fragmentation, thereby increasing the risk of adverse events for patients with MLTC (DuGoff et al., 2016). Similarly, well-designed care coordination programmes have demonstrated improvements in the quality of care and patients' well-being (Peikes et al., 2009), whereas poor care coordination can result in being referred across multiple hospitals, which delays the commencement of treatment or the prescription of medications (Thompson et al., 2025). This indicates the need to redesign MLTC healthcare services that prioritise both continuity of care and care coordination.

## 3. Methods
### 3.1. Data collection

This study aims to explore and identify priority healthcare service problem areas for MLTC from real-life experiences and stories of relevant patients and carers. To ensure MLTC coverage, four datasets relevant to different areas of MLTC are extracted from the Care Opinion platform by using the APIs provided by the platform, adopting corresponding keywords as indicated in Table 1. Care Opinion is a non-profit organisation funded in the UK, allowing people to share their good and bad experiences and stories of health and care services. The data extracted was limited to the past 5 years (2018-2023) at the time of extraction to reflect the recent health and care service experiences. Dataset 1 "Beyond MLTC" includes experiences and stories related to MLTC that require coordination and continuity. Dataset 2 "Frailty" contains stories related to frailty. Dataset 3 "Children" involves stories of children with long-term complex conditions, including mental health issues. Dataset 4 "MLTC" contains stories not related to frailty or children.

Table 1. Four datasets for MLTC

| Datasets | Keywords |
|---|---|
| Beyond MLTC | "care co-ordinat*", care co-ordination, care coordinator, coordination of care, "co-ordination of care", Carer, Carers, "co-ordination of staff", coordination of staff, "staff co-ordination*", coordination of services, "care continuity", "care conti*", "continuity of care", "continuous care", Coordination, "Co-ordinator", "hospital to home", "information shar*", information sharing, "sharing information" |
| Frailty | frail, frailty, frail elderly, "frail*" |
| Children | "child*" with diabetes/arthritis/cerebral palsy, "carer*" with diabetes/arthritis/cerebral palsy, "parent*" with diabetes/arthritis/cerebral palsy, "juvenile*" with arthritis, "child*" with mental illness, "mental health*" with child/children/childcare/child care, "carer*" with mental illness, "mental health*" with carer/carers, "parent*" with mental illness, "mental health*" with parent/parents/parenting. |
| MLTC | "multiple condition*", "multiple long term condition*", long term condition. |

The datasets extracted were subsequently scanned and cleaned, such as removing stories that are not relevant to the datasets, and formatted into Excel sheets for the following coding and analysis. After cleaning, a total of 2320 real-life stories of MLTC patients' and their carers' experiences of healthcare services are captured. This involves 1447 stories in the "Beyond MLTC" dataset, 727 stories in the "Frailty" dataset, 97 stories in the "Children" dataset, and 49 stories in the "MLTC" dataset.

## 3.2. Coding

Thematic coding was applied to identify recurring MLTC healthcare service experience themes. A coding scheme was developed by a group of researchers using a hybrid (deductive-inductive) approach. Drawing on existing literature that emphasises the central role of continuity of care and care coordination in MLTC services, these were incorporated as a priori core themes within the coding framework. Additional categories and subthemes emerged inductively through analysis of the stories. An initial set of 150 stories was double-coded by two researchers to assess inter-rater reliability, with disagreements resolved through discussion, to ensure the coding scheme and results are reliable. The refined scheme was then applied across the dataset. As shown in Table 2, the coding scheme involves eight categories (Level 1 Codes), with five categories further refined into sub-level categories (Level 2 Codes).

To code each story, the coder first identified whether the experience described was of a positive or negative nature, and then applied the relevant Level 1 and Level 2 codes. For example, if a story was coded as "Positive", "Staff" and "Continuity of care - informational", this indicates a positive experience about staff and informational continuity of care in this story. In those cases that contained both positive and negative experiences, the story was split and coded separately for each aspect. This is to ensure that the coding is mutually exclusive, with only unidirectional sentiments attached to each coded segment.

Table 2. Explanations of the coding scheme

| Codes - Level 1 | Codes - Level 2 | Explanations |
| --- | --- | --- |
| Sentiment - Positive | \ | Positive experience |
| Sentiment - Negative | \ | Negative experience |
| Staff | \ | Experiences related to the performance or behaviour of staff |
| Communication | Family/Caregiver | Communications between family/caregiver and the healthcare- hospital/clinic/GP/doctor/nurse/ etc. |
| | Internal | Communications within healthcare providers, such as between doctors and nurses |
| | Patient | Communications between the patients and the healthcare providers (hospital/clinic/GP/doctor/nurse/ etc.) |
| | Unclear | Experience relates to communication; however, the nature is ambiguous and cannot be classified as one of the above |
| Facility/Infrastructure | Medical equipment | Performance/quality of medical equipment, such as health monitors |
| | Safety-related | Performance/quality of safety-related infrastructure, such as handrails for frail people |

| | Other | Performance/quality of other equipment/infrastructure, such as the water machine and background music |
|---|---|---|
| Temporal | Patient transition | Waiting time once the patient has been accepted or is under care |
| | Access to services | Access-related issues, such as waiting for appointments, or issues relating to travelling to the facility |
| Continuity of care | Informational | The use of information on past events and personal circumstances to make current care appropriate for each individual |
| | Management/Longitudinal | A consistent and coherent approach to the management of a health condition that is responsive to a patient's changing needs |
| | Relational | An ongoing therapeutic relationship between a patient and one or more providers |
| | Unclear | The context of continuity is not specified (may relate to any subtype) |
| Care coordination | Internal | Coordination among teams within the same healthcare organisation (e.g., within a hospital) |
| | External | Coordination between organisations, such as between the hospital and an external service provider |
| | Unclear | The coordination context is unclear and may be internal or external. |

## 4. Results

For the positive MLTC stories identified from the four datasets, the quantitative results are presented in Table 3, reporting the number of stories coded according to the coding scheme developed. Overall, there are 1408 positive stories identified, including 957 positive stories from the "Beyond MLTC" dataset, 372 stories from the "Frailty" dataset, 53 from the "Children" dataset, and 26 from the "MLTC" dataset. In the "Beyond MLTC" dataset, there are 840 stories containing positive aspects related to "Staff", 483 related to "Communication", 59 related to "Facility/Infrastructure", 160 related to "Temporal", 418 related to "Continuity of care", and 332 related to "Care coordination". Please note that individual stories may span multiple Level 1 codes, such as "Staff", "Communication", and "Continuity of care", and that a Level 1 code may in turn encompass multiple Level 2 codes. For example, the story contains the "Continuity of care" aspect may further include both "Informational" and "Relational" aspects. For the "Frailty" dataset, there are 341 stories related to "Staff", 144 stories related to "Communication", 28 related to "Facility/Infrastructure", 55 related to "Temporal", 58 related to "Continuity of care", and 86 related to "Care coordination". For the

"Children" dataset, 49 stories are related to "Staff", 17 stories are related to "Communication", 5 are related to "Facility/Infrastructure", 11 are related to "Temporal", 25 are related to "Continuity of care", and 13 are related to "Care coordination". For the "MLTC", there are 21 related to "Staff", 7 related to "Communication", 5 related to "Facility/Infrastructure", 7 related to "Temporal", 10 related to "Continuity of care", and 3 related to "Care coordination".

Table 3. Positive MLTC stories

| Positive | Staff | Communication | | | | Facility | | | Temporal | | Continuity | | | | Coordination | | |
|---|---|---|---|---|---|---|---|---|---|---|---|---|---|---|---|---|---|
| | | c1 | c2 | c3 | c4 | f1 | f2 | f3 | t1 | t2 | n1 | n2 | n3 | n4 | o1 | o2 | o3 |
| Beyond | 840 | 123 | 20 | 411 | 3 | 0 | 2 | 57 | 74 | 107 | 61 | 320 | 172 | 0 | 93 | 78 | 168 |
| Frailty | 341 | 63 | 7 | 91 | 22 | 1 | 1 | 26 | 21 | 36 | 10 | 51 | 7 | 1 | 38 | 35 | 23 |
| Children | 49 | 12 | 0 | 8 | 0 | 0 | 0 | 5 | 5 | 7 | 2 | 16 | 9 | 2 | 3 | 1 | 9 |
| MLTC | 21 | 2 | 1 | 5 | 0 | 0 | 0 | 5 | 1 | 6 | 2 | 8 | 2 | 0 | 0 | 0 | 3 |

* c1 - Family/Caregiver, c2 - Internal, c3 - Patient, c4 - Unclear; f1 - Medical equipment, f2 - Safety-related, f3 - Other; t1 - Patient transition, t2 - Access to services; n1 - Informational, n2 - Management/Longitudinal, n3 - Relational, n4 - Unclear; o1 - Internal, o2 - External, o3 - Unclear.

For the negative MLTC stories, the quantitative results are shown in Table 4, with a total of 912 negative stories. This includes 490 "Beyond MLTC" negative stories, of which 168 stories are related to "Staff", 307 stories are related to "Communication", 80 stories are related to "Facility/Infrastructure", 212 stories are related to "Temporal", 328 stories are related to "Continuity of care", and 99 stories are related to "Care coordination". There are 355 negative "Frailty" stories, including 131 stories related to "Staff", 162 stories related to "Communication", 109 related to "Facility/Infrastructure", 166 stories related to "Temporal", 43 stories related to "Continuity of care", and 35 stories related to "Care coordination". For the 44 negative "Children" stories, 22 related to "Staff", 11 related to "Communication", 10 related to "Facility/Infrastructure", 17 related to "Temporal", 12 related to "Continuity of care", and 2 related to "Care coordination". Among the 23 MLTC negative stories, there are 7, 8, 3, 13, 10, and 4 stories containing aspects related to "Staff", "Communication", "Facility/Infrastructure", "Temporal", "Continuity of care", and "Care coordination", respectively.

Table 4. Negative MLTC stories

| Negative | Staff | Communication | | | | Facility | | | Temporal | | Continuity | | | | Coordination | | |
|---|---|---|---|---|---|---|---|---|---|---|---|---|---|---|---|---|---|
| | | c1 | c2 | c3 | c4 | f1 | f2 | f3 | t1 | t2 | n1 | n2 | n3 | n4 | o1 | o2 | o3 |

| | | | | | | | | | | | | | | | | |
|---|---|---|---|---|---|---|---|---|---|---|---|---|---|---|---|---|
| Beyond | 168 | 69 | 38 | 23 | 2 | 5 | 18 | 63 | 140 | 81 | 149 | 242 | 119 | 2 | 7 | 10 | 83 |
| Frailty | 131 | 96 | 13 | 51 | 21 | 7 | 21 | 92 | 82 | 93 | 17 | 22 | 2 | 7 | 8 | 15 | 12 |
| Children | 22 | 7 | 0 | 5 | 0 | 1 | 1 | 9 | 6 | 12 | 6 | 2 | 2 | 3 | 0 | 0 | 2 |
| MLTC | 7 | 1 | 0 | 7 | 0 | 0 | 0 | 3 | 4 | 9 | 4 | 7 | 1 | 0 | 1 | 0 | 3 |

*\* c1 - Family/Caregiver, c2 - Internal, c3 - Patient, c4 - Unclear; f1 - Medical equipment, f2 - Safety-related, f3 - Other; t1 - Patient transition, t2 - Access to services; n1 - Informational, n2- Management/Longitudinal, n3 - Relational, n4 - Unclear; o1 - Internal, o2 - External, o3 - Unclear.*

To yield further insights from the coding results, qualitative analysis is performed to look at each code in more detail. Common dimensions of each code for both positive and negative stories across the four datasets are extracted by looking at the corresponding stories in detail. Summaries of the common dimensions identified for positive and negative stories are presented in Tables 5 and 6, respectively, with relevant sample example stories provided. For example, in Table 5, according to the positive stories in "Communication", the common dimension identified is "effective and patient". This shows that multiple long-term condition patients/carers report positive communication experiences when communication is effective and patient.

Table 5. Common dimensions in positive stories

| Code | Dimensions | Examples |
|---|---|---|
| Staff | Kind, helpful, and reliable | "*the staff are really lovely*", "*they have been a great help*", "*everyone is helpful and direct*", "*nurses are so good*", "*staff are amazing*", "*staff have been wonderful*", "*staff are great*", "*very reliable*" |
| Communication | Effective, patient | "*took time to speak with my mum*", "*plenty of opportunity to ask questions*", "*receive call back for advice*", "*telephoned to sort out problems*", "*careful explanations*" |
| Facility | Good facility, safe, clean | "*private room looking out onto beautiful gardens*", "*look at equipment… make sure it was the right height and in the right place…advice on trip hazard*", "*facilities are good…TV…*", "*facilities are all fine*", "*ward kept clean*" |
| Temporal | Swift, prompt, efficient | "*in touch with me the day I came out*", "*provided swift treatment*", "*promptly attended by doctors and nurses*", "*…within 2 hours of the referral*", "*taken straight in…back out before appointment time*" |

| Code | Dimensions | Examples |
| --- | --- | --- |
| Continuity of care | Reliable follow-up, long-term support | "*follow up call from nurse… physio to give further support*", "*excellent care and attention for many years*", "*sent out nurses…good check over*", "*visited at home…visited to further assess…further visits from the team…*" |
| Care coordination | Coordinated, well-managed | "*went to hospital… handed over to virtual ward… managing condition at home*", "*discussing between receptionist, pharmacist and GP*", "*reception staff, practice nurse, and pharmacist has telephoned*", "*has carers, social worker and care team*" |

Table 6. Common dimensions in negative stories

| Code | Dimensions | Examples |
| --- | --- | --- |
| Staff | Negligent care, rude | "*refuse care to patient*", "*rude*", "*rude and couldn't care less*", "*gp didn't take any control*", "*need to understand they have a duty of care…*", "*nurses chatting… appearing disinterested*", "*refuse to consult…*" |
| Communication | Poor explanation, unhelpful responses | "*there needs to be a lot more communication*", "*multiple attempts at contacting…but unable to help*", "*phoned my landline which I repeatedly asked not to*", "*explained to nurses, but none knew my disease*", "*hostile and defensive in their responses*" |
| Facility | Inadequate equipment, uncomfortable | "*electricals plugged haven't been pat tested*", "*pop and rock music is stressful in waiting room*", "*monitors for checking glucose and ketone levels are poor*", "*hospital mattress not good*" |
| Temporal | Delays, long waits, and access barriers | "*after the very long wait in A&E*", "*no idea how to get an appointment… called and waited in a queue for 1 hour 20 minutes…*", "*attended appointment promptly…a queue of at least 20 people*" |
| Continuity of care | Lack of continuity and follow-up support | "*cannot see the same doctor again next time*", "*discharged to an advisory team that seems non existent*", "*report written have no details…no evidence to get support in school*", "*waiting for treatment… there is no record…*", "*circular motion of referral, discharge, urine samples to rule out UTI, for three times*" |
| Care coordination | Lack of coordinated care, misaligned processes. | "*made 3 referrals but still not had one*", "*hospital chase up for my patient record from pre-op clinic*", "*two hospitals involved… neither arranged a care plan*", "*did say do an Xray… but arranged transport home*", "*Frailty team told I could go home…ward staff not aware of my discharge*" |

## 5. Discussion

It was initially expected that open submissions would generate a diverse mix of patient experiences. However, it became evident, even within the initial 200 stories, that the majority of views expressed were polarised in nature. The dominant group represented the patients who either shared a positive or a negative experience. Mixed experiences, that combined both positive and negative aspects, were very rare, and entries expressing moderate or balanced views were virtually absent. This aligns with an increasingly recognised pattern in research on online reviews, often described as the "J-shaped" distribution (Hu et al., 2009), where a large volume of positive reviews is followed by a sharp dip in neutral opinions and increased concentration of negative feedback. Given the outsourced, voluntary nature of care opinion submission in the healthcare context, our observation appears to be mirroring this broad trend.

While "Staff" and "Communication" were the most prevalent themes across positive stories in all datasets, an asymmetry became evident when comparing positive and negative narratives. Positive experiences were largely driven by interpersonal factors, where patients and carers frequently described kindness, empathy and effective communication as defining features of good care. In contrast, negative experiences were dominated by systemic issues such as "Continuity of care", "Care coordination", and "Temporal - Access to services", which emerged as the most significant areas of challenge for MLTC patients and carers. These themes were associated with the highest numbers of negative stories across datasets.

It should be noted that, as Continuity of care" and "Care coordination" were included among the initial search terms, their prevalence in the results partly reflects the scope of the dataset. Nevertheless, their consistent recurrence across datasets and in both positive and negative contexts reinforces their central importance in shaping MLTC care experiences. The following sections, therefore, focus on these themes not only due to their frequency, but also because they represent enduring and multidimensional service challenges confirmed through qualitative analysis.

"Continuity of care" was one of the most frequently coded themes in both positive and negative stories. Patients and carers valued reliable follow-up and long-term support, describing excellent care and attention over many years, proactive home visits, and regular check-ins from staff as examples of good practice. In contrast, negative stories revealed gaps in continuity and follow-up support, including poor handovers, missing information, the inability to see the same clinician, and repeated examinations. For MLTC patients, who require continuous care and depend on a complete understanding of their history, failures in continuity impose a heavy burden and increase risk.

"Care coordination" followed a similar pattern. When coordination worked well, patients and carers described the care as well-managed and coordinated, with smooth transitions from hospital to home, effective collaboration between GPs, pharmacists, and nurses, and active involvement of carers and social workers. When coordination failed, poor management often led to severe disruptions, such as unacted referrals and conflicting messages from different healthcare teams and providers. As MLTC care often involves and spans multiple healthcare providers, strong coordination is essential to prevent confusion and delays.

"Temporal", particularly "Access to services", also played a critical role in both positive and negative MLTC patients' and carers' experiences. Positive stories described swift, prompt, efficient access to healthcare services, showing how smoothly and quickly the care was delivered. Negative stories were often linked to long waits and access issues, including chaotic booking systems, long waits and access issues, limited transport or home-visit options, and delays once already in care. For MLTC patients, such barriers may further harm health conditions and create extra stress. However, delays associated with "Patient transition" represent a widespread challenge within healthcare services and are not specific to the experiences of MLTC patients and carers. Therefore, "Temporal - Patient transition" is not considered a priority challenge for MLTC.

Although "Staff" and "Communication" were often coded in both negative and positive stories, they were not identified as the priority challenges for MLTC. One reason is that they were coded much more frequently in positive stories than in negative ones. Patients and carers consistently appreciated "Staff" for being kind, helpful, and reliable, and valued "Communication" that was effective and patient. Even when negative stories indicate negligent and rude "Staff" and poor and unhelpful "Communication", these are less frequent or less severe compared with "Continuity of care", "Care coordination" and "Temporal". Moreover, many negative perspectives coded under "Staff" and "Communication" were in fact underlying factors for causing failures in "Continuity of care", "Care coordination" and "Temporal". Another reason is that these staff behaviours and communication, while important, are common across many areas of healthcare and are not a unique characteristic of MLTC-related care experience. Additionally, while issues such as "Continuity of care" and "Care coordination" are also widespread, they are likely to have particularly severe consequences for patients with MLTC, whose care depends on consistent information sharing, long-term management, and collaboration across multiple providers.

"Facility/Infrastructure" was also not considered a priority area. It constituted a smaller proportion of negative stories, while only a few negative stories were related to medical equipment and safety-related issues, which tended to be site-specific rather than systemic.

To explore the three priority challenges further, the negative stories of the three problem areas at Level 2 Codes are analysed qualitatively to extract the common dimensions. This includes "Continuity of care - Informational", "Continuity of care - Management/Longitudinal", "Continuity of care - Relational", "Care coordination - Internal", "Care coordination - External", and "Temporal - Access to services". A summary of the further analysis is presented in Table 7.

Table 7. Common dimensions in negative stories - further analysis

| Code | Dimensions | Examples |
|---|---|---|
| Continuity of care - Informational | Information not passed on | "*got called backed by a lady don't know what happened*", "*test I have no knowledge of is on my records*", "*waiting for treatment… there is no record…*", "*circular motion of referral, discharge, urine samples to rule out UTI, for three times*" |
| Continuity of care - Management/Longitudinal | Poor handover and aftercare | "*report written have no details…no evidence to get support in school*", "*staff visited…took test…no more care or contact*", "*no after care administered*", "*promised a call…no call received…process start again*", "*discharged to an advisory team that seems non existent*" |
| Continuity of care - Relational | No relationship between patients and care providers | "*cannot see the same doctor again next time*", "*difficult to see doctors…need to explain complicated history*" |
| Care coordination - Internal | Miscommunication, lack of oversight | "*calls being pooled for answering by both sites…didn't check which site I required*", "*did say do an Xray… but arranged transport home*", "*Frailty team told I could go home…ward staff not aware of my discharge*", "*CT delay…waiting for the staff member to come back from…*", "*seen by a nurse…ask to wait…left unattended*" |
| Care coordination - External | Lack of joint care | "*made 3 referrals but still not had one*", "*hospital chase up for my patient record from pre-op clinic*", "*two hospitals involved… neither arranged a care plan*", "*transport is booked and can't be changed*", "*discharged…no carer arrangements made*", "*…had staples in…no one contacted GP or nurse to remove these*" |
| Temporal - Access to services | Access barriers - physical and digital | "*without transport…difficult to get to…and I have disability*", "*fell at home…asked for a home visit…did not want to visit*", "*directed to booking an appointment online…there weren't any*", "*the appointment system is chaotic…no matter when you call…appointment gone*", "*telephone to make an appointment...told to call back…told to visit in person to organise the appointment…can barely walk*" |

For "Continuity of care", three types of continuity failures are highlighted. When "Informational" continuity fails, information, such as records and test results, is not passed on, and patients are often left confused and experience delays in care. When "Management/Longitudinal" continuity fails due to poor handover and aftercare, it becomes challenging for patients to seek follow-up care and treatment. When "Relational" fails due to weak ongoing relationships between patients and providers, the absence of a consistent clinician leads to delays and unnecessary repetitions. All three continuity issues are associated with the same underlying factor that the healthcare service system lacks reliable transfer and use of information across providers and over time (Crooks & Agarwal, 2008). Therefore, this raises the first healthcare service design direction for MLTC:

**Direction 1**: How can relevant information about an MLTC patient's records and results be transferred and used reliably over time with the corresponding care providers to ensure continuity of care?

For "Care coordination", both internal and external issues were observed in patients' and carers' experiences. This includes internal miscommunication and lack of oversight, as well as external failures where care providers did not communicate effectively with each other or with other services to act jointly. In line with findings in care coordination practice (Diane Shannon, 2012; Jones et al., 2015; Thompson et al., 2025), poor communication between and within hospitalists and care providers emerges as the most significant barrier to achieving coordinated care. Such communication gaps often create confusion for patients and carers, cause contradictory actions across services, and lead to unnecessary delays in treatment and support. This highlights the need for improved communication approaches across internal teams and external providers, leading to the second healthcare service design direction for MLTC:

**Direction 2**: How can internal hospital teams and external care providers communicate effectively (between and within) and act jointly to avoid confusion, contradictions and delays to ensure coordinated care?

For "Temporal - Access to services", a range of access barriers were mentioned by MLTC patients and carers, causing challenges and difficulties for them in accessing healthcare services when needed. These include physical barriers, such as difficulty with transport or mobility issues to attend care treatments, and digital barriers, such as non-functioning booking systems that cause obstacles to making appointments. Similar concerns have been raised in broader healthcare contexts, where limited and inconvenient access to healthcare services and facilities has been evidenced to cause failures in various healthcare programmes (Patel et al., 2024). These experiences showed how current healthcare access pathways are not designed around the complex health needs of MLTC patients, which creates health risks and stress. This informs the third healthcare service design direction for MLTC:

**Direction 3**: How can healthcare services be organised to ensure inclusive access for MLTC patients, with minimal physical (e.g. transport limitations, mobility challenges) and digital barriers (e.g. booking constraints)?

## 6. Conclusions

This research proposes the use of data to identify what constitutes positive or negative user experiences within healthcare in the UK. A total of 2,320 real-life patient and carer experiences were analysed to inform future directions for designing healthcare services, particularly for people with more than one long-term condition. Although many stories highlighted the positive aspects of current MLTC healthcare services, most notably the kindness, empathy, and professionalism of staff and the quality of communication, "Continuity of care", "Care coordination", and "Temporal - Access to services" were identified as the priority challenges of MLTC healthcare services. Further qualitative analysis informed three healthcare service design directions to improve the experiences of MLTC patients and carers. Specifically, (1) to ensure continuity of care, (2) to deliver coordinated care, and (3) to provide inclusive access to care. This paper demonstrates how real-life patient and carer stories and experiences data can be leveraged to inform healthcare service design directions, balancing the richness of qualitative data with the representativeness of large-scale patient experience datasets to provide a robust foundation for identifying systemic challenges. However, as the Care Opinions platform relies upon voluntary contributions, the stories and experiences shared tend to over-represent extreme positive or negative experiences, with fewer moderate accounts. Future research will include interviews and focus group studies to capture a broader range of perspectives, providing deeper insights into the challenges and needs of MLTC patients and carers.

**Acknowledgement**

This work is funded by SPHERE: Systems and People-Centric Innovation in Healthcare Redesign (NIHR157673). We acknowledge Nav Mustafee for his role in acquiring the datasets.